\documentclass[10pt,conference]{IEEEtran}
\usepackage[utf8]{inputenc}
\usepackage{cite}
\usepackage{url}
\usepackage{hyperref}
\usepackage{comment}
\usepackage[normalem]{ulem}
\usepackage[pdftex]{graphicx}
\graphicspath{{./fig/}}
\DeclareGraphicsExtensions{.pdf,.jpeg,.png}
\usepackage[caption=false,font=footnotesize]{subfig}
\usepackage{fixltx2e}
\usepackage{pgfplots}

\begin{document}

\title{Traffic Management Applications \\ for Stateful SDN Data Plane}

\author{
  \IEEEauthorblockN{
    Carmelo Cascone\IEEEauthorrefmark{1}\IEEEauthorrefmark{2},
    Luca Pollini\IEEEauthorrefmark{3}, 
    Davide Sanvito\IEEEauthorrefmark{3},
    Antonio Capone\IEEEauthorrefmark{1}
  }
  \IEEEauthorblockA{
    \IEEEauthorrefmark{1}
    Dipartimento di Elettronica, Informazione e Bioingegneria, Politecnico di Milano, Italy\\
    Email: antonio.capone@polimi.it,
  }
  \IEEEauthorblockA{
    \IEEEauthorrefmark{2}
    D\'epartement de g\'enie \'electrique, \'Ecole Polytechnique de Montr\'eal, Canada\\
    Email: carmelo.cascone@polymtl.ca
  }
  \IEEEauthorblockA{
    \IEEEauthorrefmark{3}
    CNIT, Consorzio Nazionale Interuniversitario per le Telecomunicazioni,  Italy\\
    Email: luca.pollini@mail.polimi.it,
    davide2.sanvito@mail.polimi.it
  }
}

\maketitle

\begin{abstract}
The successful OpenFlow approach to Software Defined Networking (SDN) allows network programmability through a central controller able to orchestrate a set of dumb switches. However, the simple match/action abstraction of OpenFlow switches constrains the evolution of the forwarding rules to be fully managed by the controller. This can be particularly limiting for a number of applications that are affected by the delay of the slow control path, like traffic management applications. Some recent proposals are pushing toward an evolution of the OpenFlow abstraction to enable the evolution of forwarding policies directly in the data plane based on state machines and local events. In this paper, we present two traffic management applications that exploit a stateful data plane and their prototype implementation based on OpenState, an OpenFlow evolution that we recently proposed.
\end{abstract}

\section{Introduction}
\label{sec:introduction}

The main innovation of SDN is the separation between control and data plane. With OpenFlow this separation is physically implemented with dumb switches processing tables of match/action rules (flow entries) instantiated by smart controllers. The controller can dynamically update forwarding policies by modifying flow entries in the switches in order to react to events that are typically notified by the switches themselves. This approach has the advantage of allowing simple network programming paradigms based on an abstraction of the network as a single entity (big switch) and application logic based on system level events and global state evolution. 

However, the logically centralized approach of OpenFlow introduces, in the best case, an additional processing delay and extra signaling. In the worst case, the use of the control path through the controller is too slow and prevents the support of network functions that need real time reactions to events. A relevant example of applications affected by the limitations of the slow control reaction of OpenFlow are those for traffic management where fast network adaptability to changing conditions is often important and events characterizing changes are usually local to the switch or data path whose forwarding behavior needs to be modified.

A few recent proposals are pushing for an evolution of the OpenFlow abstraction that allows to introduce adaptation of the forwarding rules based on local events observed by the switch \cite{Bos14,Mos14}. OpenState \cite{bianchi14} is an evolution of the OpenFlow abstraction, proposed by some of the authors, that has the remarkable advantage of defining a stateful data plane with minimal modifications to OpenFlow. OpenState retains the OpenFlow property of a centralized control logic and delegates the application of different sets of pre-instantiated forwarding rules to switches according to local states. Local events that can trigger state transitions are packet arrivals, measurements and timers.

In this paper we present two traffic management applications, namely forwarding consistency and failure recovery, that benefit from a stateful data plane. Both applications can also be functional blocks of more complex SDN applications for traffic engineering and resource management.

Forwarding consistency is required in all scenarios where some type of load balancing between different links/path is adopted but consistent forwarding on the same output port for packets of the same session must be guaranteed. The definition of session depends on the scenario and can go from micro-flows at the IP layer to bursts of packet transmissions of a transport connection. In OpenFlow, the ``select'' group entry allows load balancing between output ports and forwarding consistency can only rely on switch specific functions external to OpenFlow (such as hashes on packet header for random port selection). No fine-grained control on the session definition can be provided. With OpenState, we show that forwarding consistency can be fully controlled by application developers based only on needs using states in the data plane in a efficient and scalable way.

Failure resilience is a fundamental requirement in any network. In OpenFlow, another group table capability named ``fast-failover'' allows a programmer to specify alternative ports to be used in case of failure. In all other cases, where backup paths are not local detours from the node that detects the failure, the network controller must be notified in order to establish a backup path by updating flow tables (e.g. path protection scenarios). This introduces signaling overhead and a recovery delay leading to possible losses. Moreover, if the controller is not available, the network cannot restore working conditions. We show that with OpenState it is possible to design a protection scheme able to recover also from non-local failures without the controller direct involvement. This allows to get fast recovery times and to overcome issues with controller unresponsiveness (high control path delay) or unreachability (controller failure).

The remainder of the paper is as follows. In Section \ref{sec:openstate} we first present an overview of OpenState and discuss related work. We then introduce our applications for traffic management in Section \ref{sec:applications}. Some numerical results are presented in Section \ref{sec:results}. Section \ref{sec:conclusion} concludes the paper.  

\section{OpenState}
\label{sec:openstate}

OpenState, proposed by some of the authors \cite{bianchi14}, is an OpenFlow extension that introduces the idea of offloading some control functions inside the devices, while still keeping the central SDN controller informed and in full control of all delegated operations and possible exceptions. The motivation beside OpenState is to prevent the controller from handling simple control tasks that require only switch-local knowledge, making it responsible only for decisions requiring network-wide knowledge. OpenState provides the ability to configure custom states inside the switch and to program how states should evolve as a consequence of packet arrivals according to an abstraction equivalent to a Mealy machine. OpenState has been implemented as an OpenFlow 1.3 Experimenter extension. The complete protocol specification along with a switch and controller implementation is available at \cite{openstatehomepage}. Finally, OpenState's hardware viability was addressed in \cite{pontarelli15}.

\begin{figure}
  \centering
  \includegraphics[width=\columnwidth]{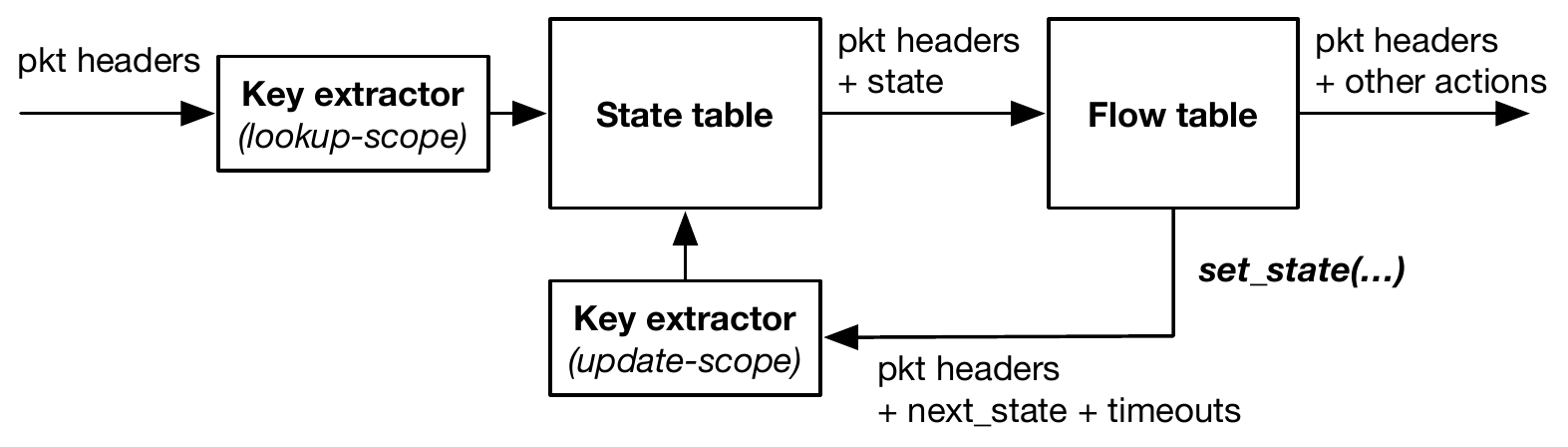}
  \caption{Simplified packet flow in OpenState}
  \label{fig:os-stateful-stage}
  \vspace{-5mm}
\end{figure}

In the OpenState pipeline, flow tables can be optionally preceded by a state table (Fig.~\ref{fig:os-stateful-stage}). The latter is intended to store ``flow states'', used to map different forwarding behaviors without involving the controller whenever a new behavior is needed. Each time a new packet comes to the switch, it is firstly handled by the state table. The state table lookup phase is performed by using a set of user-specified fields described by a so called ``lookup-scope'', which purpose is to define the flow identifier (key) to match a specific entry in the state table. Upon matching an entry, the packet is returned with an associated state label equivalent to an additional header field. If a packet does not match any state entry, a 0 (default) state is returned. The packet is then sent to the flow table where the standard OpenFlow processing has been extended with the additional match field ``state'' and a new ``set-state'' action used to insert/update entries in the state table. When adding a set-state action in an OpenFlow's flow-mod, a programmer explicitly specifies the new state label to be used for future packets of the same flow. Alternatively, by defining a different ``update-scope'', it is possible to point to a different state entry than the one specified by the lookup-scope, allowing for cross-flow state updates\footnote{The immediate example is the case of a MAC learning switch where states are used to store the location (output port) of a given host. In the MAC learning scheme, packets are forwarded based on the Ethernet destination address, while, for the same packet, the location of the Ethernet source address is updated using the packet input port. This simple behavior can be modelled using $lookup\_scope=[mac\_dst]$ and $update\_scope=[mac\_src]$. The complete example of a MAC learning switch implemented with OpenState is available at \cite{bianchi14}.}. Moreover, idle and hard state timeouts can be defined and are equivalent to those used in OpenFlow flow entries. In contrast to OpenFlow, a programmer can optionally specify a rollback state (non default) to be used when a timeout expires.

\subsection*{Related works}

Recent works have tried to rethink the OpenFlow data plane abstraction \cite{Bos13},\cite{Bos14},\cite{Son13}. In \cite{Bos13}, RMT (Reconfigurable Match Tables) are introduced to make matching more flexible on arbitrary header fields and extend the action set with a programmable set of primitives. A radical solution to switch programmability limitations is proposed in \cite{Bos14}: P4. P4 is a high-level language to program packet processors which focuses on protocol-independence. A similar approach is proposed in \cite{Son13} with Protocol Oblivious Forwarding (POF) abstraction model. Similarly to OpenState, FAST \cite{Mos14}, proposes the use of state machines to modify the switches' forwarding behavior. Although, its data plane design is different and it makes use of variables and functions to define events and transitions, whose hardware implementation may be not trivial.

\section{Applications for stateful data plane}
\label{sec:applications}

\subsection{Forwarding consistency}
\label{sec:fwc}

Load balancing traffic over multiple paths (also known as load sharing) is an important feature that allows flexible and efficient allocation of network resources. The trick here is to have network switches use i) a link selection scheme that guarantees the desired (optionally weighted) splitting and, most important, ii) consistency on the forwarding of packets of the same transport layer flow (i.e. TCP) in order to avoid packet reordering at the receiver, which can cause unnecessary throughput degradation.

Starting from OpenFlow 1.1, the select group type has been introduced to support load sharing over multiple ports. Citing the latest OpenFlow 1.5 specification \emph{``Packets are processed by a single bucket in the group, based on a switch-computed selection algorithm (e.g. hash on some user-configured tuple or simple round robin). All configuration and state for the selection algorithm is external to OpenFlow.''}. Thus in OpenFlow selection and consistency are tied together and left to vendors' implementations. For example, HP OpenFlow switches use a per-packet round-robin scheduler with no consistency features \cite{hp-of13-manual}; older versions of Open vSwitch used only an hash on the Ethernet destination address (without any proper rationale behind this decision \cite{pfaff14}), while more recent versions expand the hash to L2, L3 and L4 fields \cite{seetharaman14}. As a further reference, in \cite{horman14} the authors describe an OpenFlow extension to let a programmer specify the selection method along with the fields used to provide consistency.

Different hashing schemes exists, each one with its associated trade offs \cite{Zhi00}, thus we argue that choosing a selection scheme should be separated from the granularity of the states required to provide consistency. For example it has been shown in \cite{kandula07} how providing consistency at level of TCP bursts (instead of pinning the whole flow to a specific path) guarantees more accurate load shares with hardly any out-of-order packet.

\begin{figure}
  \centering
  \includegraphics[width=0.9\columnwidth]{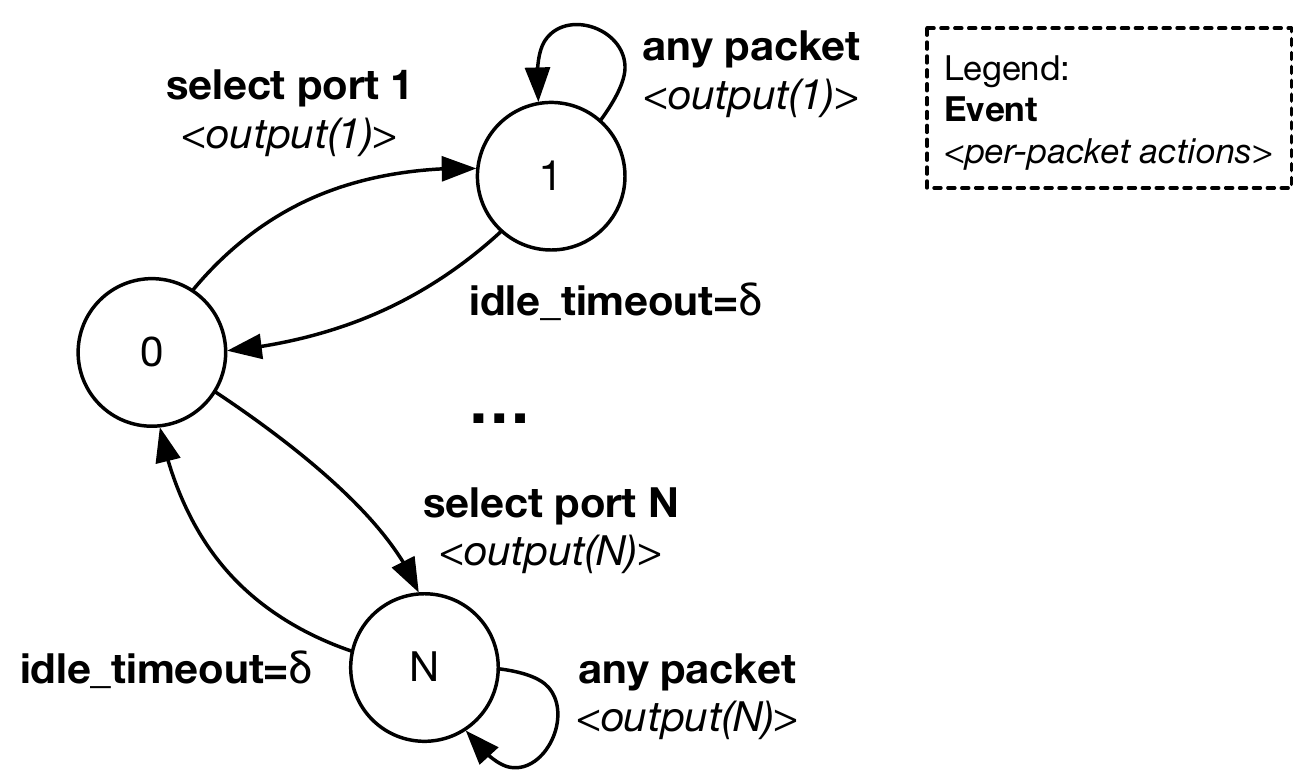}
  \caption{Mealy machine for the forwarding consistency scheme.}
  \label{fig:fw-cons-fsm}
\end{figure}

\begin{figure}[b]
  \centering
  \includegraphics[width=0.9\columnwidth]{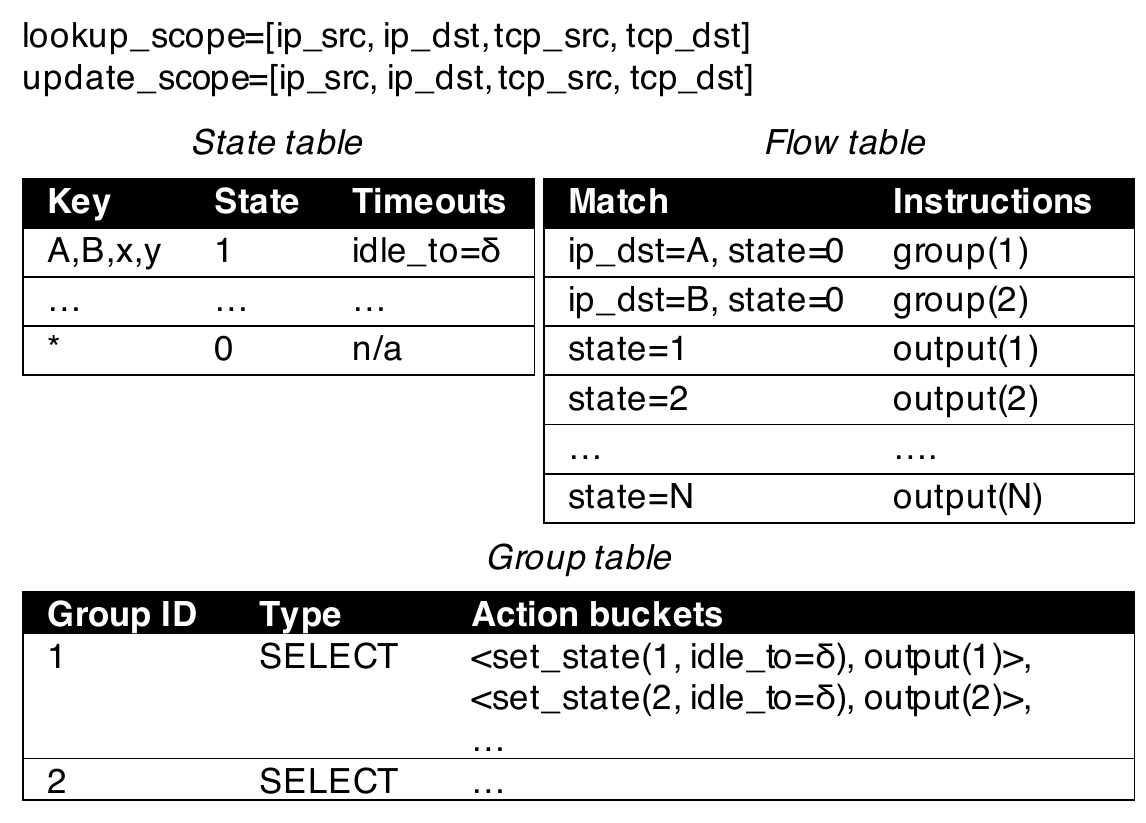}
  \caption{Example of an OpenState implementation of a destination-based load balancer using the forwarding consistency mechanism described in Fig.~\ref{fig:fw-cons-fsm}}
  \label{fig:1-t-m-table}
\end{figure}

By using flow states and associated idle timeouts, OpenState allows a programmer to choose the granularity and the lifetime of a forwarding decision. Figure \ref{fig:fw-cons-fsm} shows the behavioral model (in form of a Mealy machine) used to implement such a scheme, while Fig.~\ref{fig:1-t-m-table} presents a detailed description of the tables needed to implement a destination-based load balancer using OpenState. The granularity of the splitting is defined using the lookup-scope, in this example a 4-tuple is used to define a unique TCP flow. For each incoming packet of a new TCP connection, a state 0 (default) is returned by the state table, the corresponding group entry is invoked by the flow table based on the matched destination IP address. Finally, a random bucket is selected from the group entry, the state is updated in the state table and the packet forwarded accordingly. Subsequent packets will be forwarded using the value returned from the state table. By using an $idle\_timeout = \delta$ we can define the lifetime of the forwarding decision. For example, with $\delta=10s$ the state will be maintained only if a packet belongs to a given TCP flow (otherwise described by a different lookup-scope, e.g. UDP flow, only L2 source-destination, etc.) is seen at least once every $10s$. In this case it is safe to say that an idle interval of $10s$ represents the end of an instance of a TCP flow. As an alternative, by using the mechanism described in \cite{kandula07} smaller values of $\delta$ can be evaluated and used to distinguish bursts of the same flow. In this case, a new forwarding decision will be taken for each burst, maximizing load share accuracy while minimizing the risk of packet reorder at the receiver.

Concerning scalability, the state table is responsible for maintanining an entry for each active TCP flow/burst. Given the exact-match nature of this table (i.e. non-wildcard, always on the same fields defined by the lookup-scope), flow states in OpenState can be stored in an ordinary (cheap) RAM-based hash table with $O(1)$ access times. 

The benefits of using OpenState to implement a flexible forwarding consistency scheme are highlighted when comparing an implementation using OpenFlow switches not providing any means of forwarding consistency, like in the HP case presented above. In this case, each time the first packet of a new instance of a transport layer flow is received by the switch, and upon selecting an output port by using the group table, the switch must inform the controller of the decision, which in turns replies by installing an higher priority flow-mod that guarantees consistency by explicitly forwarding all packets of that flow using the previously selected output port. It is clear how the switch-controller RTT and the processing delay at the (logically centralized, i.e. distributed) controller make this approach hardly scalable in large networks with an increasing arrival rate of new flows. The same reactive mechanism applies when a different hashing scheme from the one implemented by switches is required. Analogously, the idea of consistently splitting packet bursts by maintain states at the controller would be totally nonviable given the high frequency of control messages needed.

Finally, we argue that a more flexible SDN/OpenFlow data plane offering load balancing features should separate selection from consistency. Vendors should be free to compete by offering different efficient selection algorithm, from simple weighted random algorithms that proportionally map packets to output ports to more advanced token counter algorithms based on feedback about past decisions (e.g. based on byte counters) \cite{kandula07}. While the granularity and lifetime of the forwarding decisions should be left to programmers, based only on the application requirements. OpenState's general-purpose stateful pipeline allows programmers to define such a behavior.

\subsection{Failure recovery}
\label{sec:ft}

Resiliency to failures (link or node) is a fundamental requirement: the ever-increasing bit rate brings to a huge amount of data traveling through the network, hence even a hundreds of milliseconds of network out-of-service implies a tremendous data loss. Different protection schemes implies different recovery delays: in the case of a link protection scheme a backup link or path towards the same downstream node is usually provisioned and allocated to serve traffic flows in case of failure of the first link. In this case, the recovery delay is almost equal to the time required to detect the failure and depends only on the detection mechanism implemented by the network device. In OpenFlow, the fast-failover group type has been introduced for this purpose, allowing a switch to handle local failures without relying on the controller and thus minimizing recovery delays and packets loss. The way the fast-failover feature works is analogous to the select group type: a programmer can define multiple action buckets for a given group entry. Each bucket is associated with an output port and only one of them is selected depending on the status (up or down) of the associated port.

\begin{figure*}
  \centering
  \includegraphics[width=\textwidth]{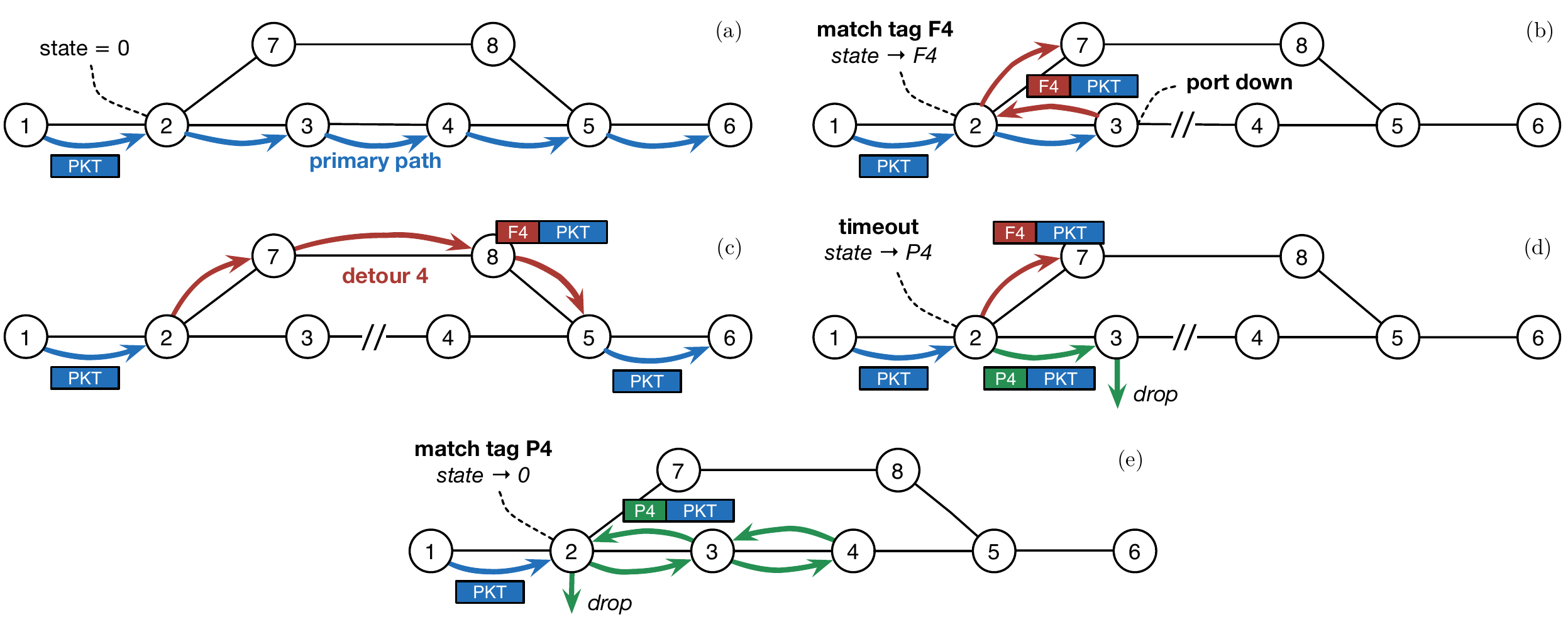}
  \caption{Failure recovery example with OpenState. The behavioral model of node 2 is shown in Fig.~\ref{fig:ft-fsm}.}
  \label{fig:ft-example}
\end{figure*}

\begin{figure}[b]
  \centering
  \includegraphics[width=0.9\columnwidth]{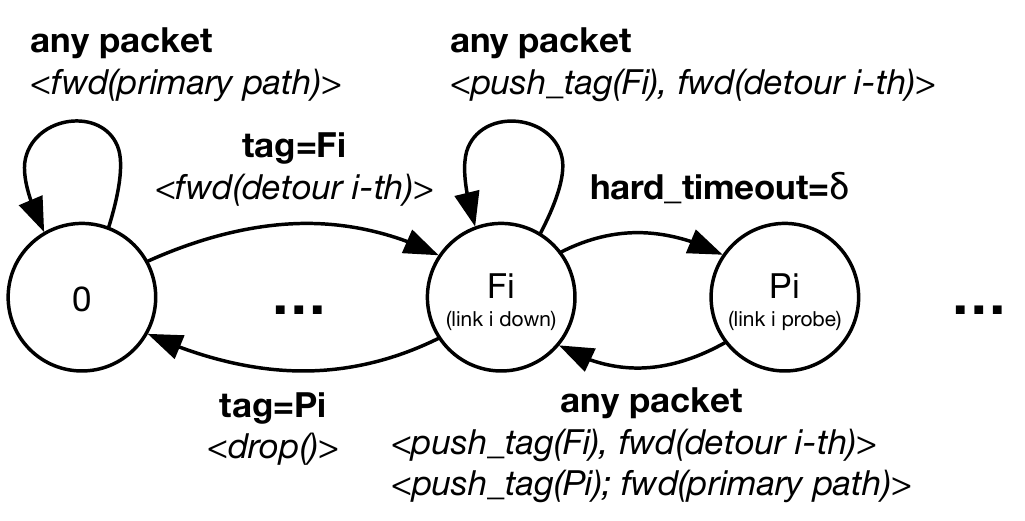}
  \caption{Failure recovery Mealy machine implemented by a reroute node.}
  \label{fig:ft-fsm}
\end{figure}

Unfortunately, it is not always the case that an alternative output port can be provisioned due to budget or topology constraints. In this case, non-local protection schemes such as path or segment protection can be used. Here, signaling is required from the node that detected the failure to one or more reroute nodes responsible to deviate traffic flows according to precomputed backup paths. In OpenFlow networks, this signaling is handled by the controller, by either receiving a ``Port Status'' notification or by periodically polling statistics from switches. Interaction with the controller often results in an overall recovery delay greater than $50ms$ \cite{staessens11}. Moreover, if the failure affects a link carrying an in-band control channel, human intervention will be probably required in order to solve the failure and re-establish connectivity with the controller.

We present here a failure recovery scheme that allows handling of non-local (distant) failures, regardless of controller reachability (i.e. completely independent from the controller, besides the initial provisioning of a backup forwarding policy). This mechanism implements the original idea proposed by some of the authors in \cite{Cap15} extended here with a probing mechanism to establish if the original failure has been resolved. Figure~\ref{fig:ft-example} shows an example of such a scheme based on OpenState. This mechanism does not require any switch-controller signaling, rather (Fig.~\ref{fig:ft-example}b) the same data packets are tagged (e.g. with a MPLS label containing the ID of the failed link) and \emph{bounced back} trough the primary path, until they reach a predetermined reroute node. Here, the match of the tagged packet in the flow table, triggers a state transition which enables (Fig.~\ref{fig:ft-example}c) the forwarding of the tagged packet and all future packets of the same flow on a preallocated detour. The tag is always maintained in the detour path (and popped when entering again the primary path, e.g. node 5 in Fig.~\ref{fig:ft-example}c) so to make detour nodes distinguish the specific forwarding to apply, allowing the allocation of the same nodes/links for different detours depending on the failed element.

Given a traffic demand, flow states are maintained only on those nodes that might act as a reroute node in case of failure. Thus, given a specific node, this will have to maintain an instance of the Mealy machine shown in Fig.~\ref{fig:ft-fsm} for each demand it has a responsibility as a reroute node, depending on the specific failure. Flow states are used to represent the state of the network. A state 0 means that the primary path is fully working, while a state F$i$ means that node $i$ is unreachable (either because of a link or node failure) and thus the demand needs to be forwarded on a failure-specific preallocated detour path. In the example of Fig.~\ref{fig:ft-example}, state F4 is used to describe the case of node 4 unreachable.

When failures are temporary (e.g. accidental disconnection of a cable in a core switch), it is important to establish the original forwarding as soon as the failure is resolved. In our scheme, the process of establishing if a failure has been resolved or not is also handled through a switch-to-switch signaling mechanism based on the same data packets. In Fig.~\ref{fig:ft-fsm} an hard timeout $\delta$ is used to periodically move from state F$i$ to a state P$i$. P$i$ is meant to serve just one packet, indeed, when in state P$i$, by matching a packet of a currently deviated demand (Fig.~\ref{fig:ft-example}d), the packet is duplicated on two ports (by means of an OpenFlow's group type ``all'') and the state is set back to F$i$. The first packet is tagged with F$i$ and sent on the detour, while the other is tagged with a special label P$i$ and forwarded on the original primary path (in Fig.~\ref{fig:ft-example}d tag P4 is used to reference a probe request for node 4). Probe packets are generated each $\delta$ interval: if the previously unreachable node receives one of them (Fig.~\ref{fig:ft-example}e), meaning that the failure has been resolved, the latter is bounced back on the primary, until it reaches the reroute node that generated it, triggering a state transition to 0 (no failures) and thus reestablishing the forwarding on the primary path.

Advantages of this scheme can be found in i) the ability of switches to autonomously and immediately react to non-local (distant) failures, independently of the controller reachability; ii) minimized packet losses due to the reuse of the same data packets to perform signalling (bounced back packets are then forwarded on the detour); iii) automated probing mechanism with a programmable trade off (based only on the $Fi$ state's hard timeout) between responsiveness and overhead (a programmer might set a very short timeout for critical links while preserving resources for others).

One might argue that bouncing back some (few) packets on the primary path and forwarding them on the detour might cause reordering at the receiver, resulting in throughput degradation equivalent to that produced by dropping those packets or relying on the slower controller intervention. In this case, the forwarding consistency scheme presented before might be integrated to distinguish between packet bursts, updating the forwarding on the reroute node only after the whole burst has been bounced back. Minimizing the risk of packet reordering by exploiting the interval between bursts.

\section{Testing scenarios and results}
\label{sec:results}

The experimental results presented in this section have been obtained using an OpenFlow 1.3 switch and controller extended to support OpenState and available at \cite{openstatehomepage}. We compare the performances of the previously presented applications with a trivial OpenFlow implementation. In our experiments, we show how handling simple control tasks at the switch with OpenState offers important gain in performances and scalability when compared to the OpenFlow reactive counterpart. One might argue that similar results might be obtained by using more advanced, specialized, and distributed controller architecture, we argue that such a choice would be more complex and expensive to manage when compared to the simplicity of the OpenState-based solution.

All tests have been performed using a Mininet VM with 4 CPU cores Intel Core i7 and 8GB of RAM available. For brevity we will refer to ``OS'' for an implementation using OpenState switches, with ``OF'' when using only OpenFlow switches.

\subsection{Forwarding consistency}

\begin{figure}[b]
  \centering
  \includegraphics[width=0.6\columnwidth]{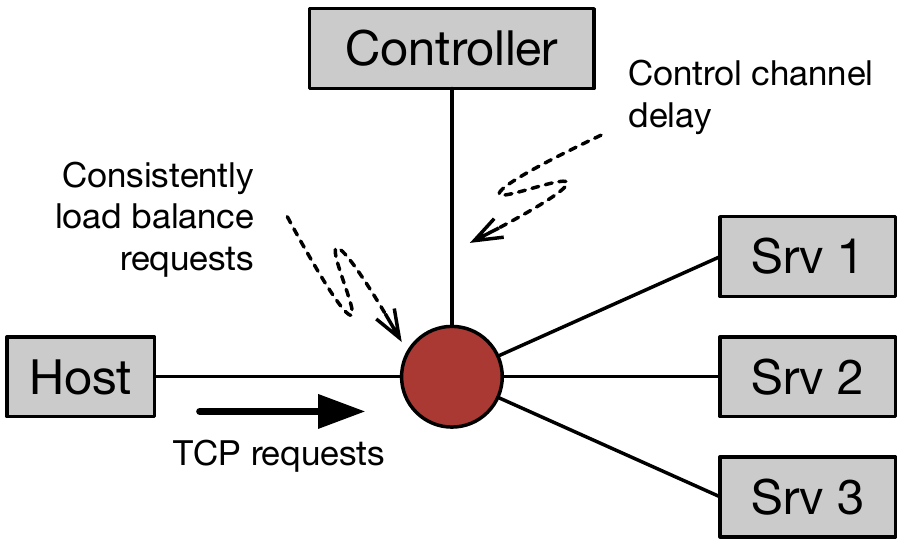}
  \caption{Topology used in the forwarding consistency experiment.}
  \label{fig:fwc_topo}
\end{figure}

To preserve computing resources, we emulated a small network with 4 hosts and a switch (Fig.~\ref{fig:fwc_topo}). One host acts as a client willing to establish TCP sessions towards a server, the switch distributes the workload across 3 replicas of the same server by consistently load balancing the incoming requests on 3 output ports. We wanted to measure the time required for the switch to ``pin'' an incoming flow to one of the possible 3 output ports. In OS we used an implementation equivalent to Fig.~\ref{fig:fw-cons-fsm}, while in OF we supposed a switch that does not guarantee consistency (as in \cite{hp-of13-manual}), for this reason the first packet of each new TCP flow is encapsulated and sent to the controller which in response randomly selects an output port and installs the corresponding flow entry in the switch, thus guaranteeing the forwarding of subsequent packets through the same port.

 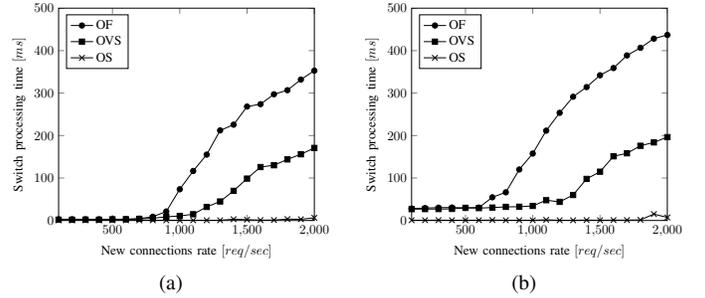
\begin{figure}
   \centering
   \subfloat[][]{
     \resizebox{0.5\columnwidth}{!}{
       \begin{tikzpicture}
         
			\begin{axis}[
				xmin = 100, xmax = 2000,
				ymin = 0, ymax = 500,
				ylabel={Switch processing time $ [ ms ] $},
				ylabel style={yshift=-0.5em},
				legend cell align=left,
				legend style ={ at={(0.03,0.97)},anchor=north west, draw=black,fill=white,align=left},
				xlabel={New connections rate $ [ req/sec ] $}
			]
			\addplot[solid, every mark/.append style={solid, fill=black}, mark=*]	
					coordinates{
				(100, 2.9377065181732185)
				(200, 2.9232831478118895)
				(300, 2.866366582565957)
				(400, 2.875729846954346)
				(500, 2.847144508361816)
				(600, 2.9094230175018305)
				(700, 4.25232653617859)
				(800, 8.802947902679445)
				(900, 21.158083033561706)
				(1000, 73.47315201759338)
				(1100, 116.5527274131775)
				(1200, 155.08306751531714)
				(1300, 212.09206552832774)
				(1400, 225.573365959617)
				(1500, 268.27973223673115)
				(1600, 273.9561419053449)
				(1700, 297.22067511487637)
				(1800, 306.82855085418214)
				(1900, 331.86329636174014)
				(2000, 352.8161724697823)
				};
			\addlegendentry{OF}
			\addplot[solid, every mark/.append style={solid, fill=black},mark=square*]                
			coordinates{
				(100, 1.7389038801193235)
				(200, 1.6358857154846191)
				(300, 1.9457373619079592)
				(400, 1.6740556001663207)
				(500, 2.7286211252212524)
				(600, 2.9475861310958864)
				(700, 3.8851903676986694)
				(800, 5.159843587875367)
				(900, 8.604983711242676)
				(1000, 10.678348803520203)
				(1100, 14.715607166290283)
				(1200, 31.863834524154672)
				(1300, 44.75702340602874)
				(1400, 69.776442694664)
				(1500, 98.61232221126556)
				(1600, 125.82339425086974)
				(1700, 130.08596727848052)
				(1800, 143.83174283504485)
				(1900, 155.903324842453)
				(2000, 170.79861195087435)
			};
			\addlegendentry{OVS}
			\addplot[solid, every mark/.append style={solid, fill=black}, mark=x, mark size=3]
			coordinates{
				(100, 0.41418278217315674)
				(200, 0.39463996887207037)
				(300, 0.40701935291290287)
				(400, 0.3945092201232909)
				(500, 0.4108429431915283)
				(600, 0.3772068738937378)
				(700, 0.3734148025512696)
				(800, 0.3781047344207763)
				(900, 0.38641200065612796)
				(1000, 0.38210585117340096)
				(1100, 0.3651430130004883)
				(1200, 0.45880076885223386)
				(1300, 0.4487101316452026)
				(1400, 2.9259679317474365)
				(1500, 1.9805556058883667)
				(1600, 0.7610444784164428)
				(1700, 1.0284677982330321)
				(1800, 3.2745146512985235)
				(1900, 2.8858459234237674)
				(2000, 6.185054969787598)
			};
			\addlegendentry{OS}
			\end{axis}
       \end{tikzpicture}
     }
     \label{fig:fwc-chart-0}%
   }
   \subfloat[][]{
     \resizebox{0.5\columnwidth}{!}{
       \begin{tikzpicture}
         \begin{axis}[ 
				xmin = 100, xmax = 2000,
				ymin = 0, ymax = 500,
				ylabel={Switch processing time $ [ ms ] $},
				ylabel style={yshift=-0.5em},
				legend cell align=left,
				legend style ={ at={(0.03,0.97)},anchor=north west, draw=black,fill=white,align=left},
				xlabel={New connections rate $ [ req/sec ] $}
			]
			\addplot[solid, every mark/.append style={solid, fill=black}, mark=*]	
					coordinates{
				(100, 27.97825138568878)
				(200, 29.288121509552003)
				(300, 30.199237132072444)
				(400, 30.478532671928406)
				(500, 29.632377266883854)
				(600, 30.421712303161616)
				(700, 54.584631681442254)
				(800, 66.29397063255308)
				(900, 120.06620993614196)
				(1000, 157.6279420375824)
				(1100, 211.47869105339052)
				(1200, 253.49883445108296)
				(1300, 291.3771611082606)
				(1400, 314.17086418585103)
				(1500, 342.03165686130524)
				(1600, 358.9864603302381)
				(1700, 388.7622078716169)
				(1800, 406.6477782539658)
				(1900, 428.29107708113145)
				(2000, 436.9978747420459)
				};
			\addlegendentry{OF}
			\addplot[solid, every mark/.append style={solid, fill=black}, mark=square*]                
			coordinates{
				(100, 26.649636387825016)
				(200, 26.813714337348937)
				(300, 26.87864689826965)
				(400, 27.25448634624481)
				(500, 29.77532205581665)
				(600, 29.147562575340274)
				(700, 30.357607865333556)
				(800, 32.05540814399719)
				(900, 32.342651939392084)
				(1000, 34.265427756309506)
				(1100, 47.708982229232795)
				(1200, 43.841422653198244)
				(1300, 60.02049169540404)
				(1400, 97.85176911354064)
				(1500, 114.88076689243316)
				(1600, 151.13342204093937)
				(1700, 158.24711902141573)
				(1800, 175.89697453975677)
				(1900, 183.8874259710312)
				(2000, 196.28634662628176)
			};
			\addlegendentry{OVS}
			\addplot[solid, every mark/.append style={solid, fill=black}, mark=x, mark size=3]
			coordinates{
				(100, 0.5804480314254761)
				(200, 0.5648705005645751)
				(300, 0.7320764303207397)
				(400, 0.5337246179580689)
				(500, 0.5725331306457518)
				(600, 0.5745407104492186)
				(700, 0.9924823999404907)
				(800, 0.5082118034362793)
				(900, 0.5481069326400757)
				(1000, 0.5490166425704955)
				(1100, 1.5603968620300293)
				(1200, 0.4776962041854859)
				(1300, 0.4620285987854005)
				(1400, 0.9727686405181883)
				(1500, 0.5175550937652588)
				(1600, 0.864972162246704)
				(1700, 0.872584843635559)
				(1800, 1.3995664596557618)
				(1900, 15.087391018867493)
				(2000, 7.108799242973326)
			};
			\addlegendentry{OS}
			\end{axis}
       \end{tikzpicture}
     }
     \label{fig:fwc-chart-12}%
   }
  \caption{Forwarding consistency results with different switch-controller RTT: (a) 0 $ms$  and (b) 12 $ms$.}
  \label{fig:fwc-charts}%
\end{figure}

Figure \ref{fig:fwc-charts} depicts the switch processing time as the connections rate increases. TCP requests are generated at an increasing rate from 100 to $2000reqs/sec$ with a step of 100, each time generating 1000 single TCP requests and repeating each experiment 10 times. For each experiment we measure the average switch delay to process a new connection, intended as the time interval between the arrival and the departure of the first TCP (SYN) packet, eventually passing from the controller in OF. In both OS and OF, the switch is based on an user space implementation \cite{ofsoftswitch13}, which offer degraded processing performances when compared to a kernel space implementation such as Open vSwitch (OVS). To offer a better term of comparison we executed the same experiments of the OF case using OVS. In Fig.~\ref{fig:fwc-chart-0} results are characterized by an almost $0ms$ switch-controller RTT, while in Fig.~\ref{fig:fwc-chart-12} an RTT of $12ms$ has been introduced to emulate an hypothetical distance between the two devices.

The results obtained show how by using a reactive OF controller approach there is a considerable increase in the switch processing time for each connection, reaching a peak of ~$400ms$ at $2000req/sec$, while in OS this value does not grow more than a few $ms$ at all tested rates. It is also noteworthy how a considerable gap from the faster OS scheme is also appreciable when using OVS. In this case, both charts show a gain in performances from the OF case thanks to the optimized packet processing offered by OVS, but still suffering from the processing and RTT required by the controller. 

\subsection{Failure recovery}

\begin{figure}
  \centering
  \includegraphics[width=0.6\columnwidth]{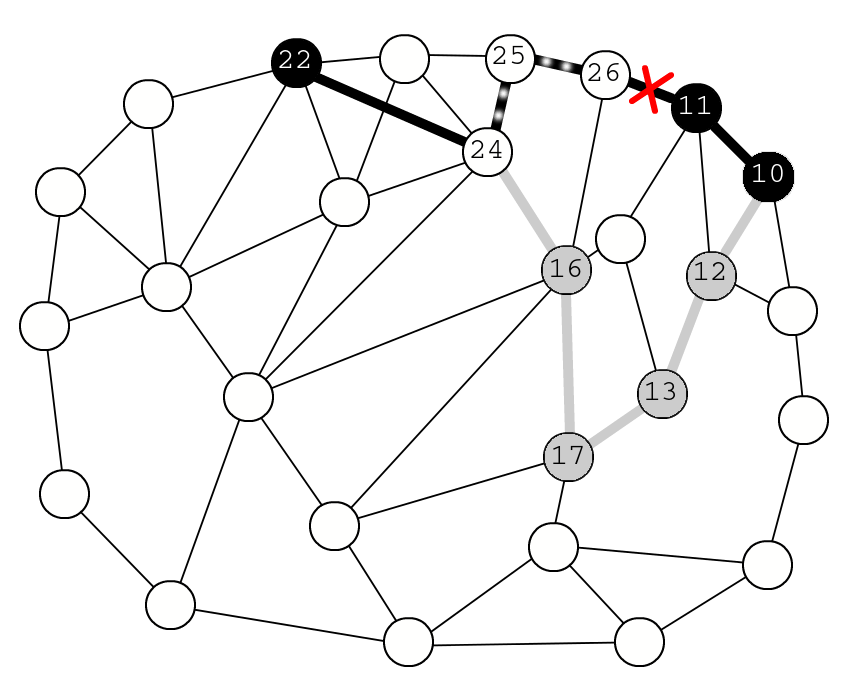}
  \caption{Topology instance used for the failure recovery experiments with failed link (11,26). In OS, when considering demand (22,10), node 26 forwards back the packet to the reroute node 24 trough the intermediate hop 25. The packet is then forwarded using the detour path 24-16-17-13-12.}
  \label{fig:failure-13-12}
\end{figure}

In order to test the OS failure recovery scheme presented in Section \ref{sec:ft}, we have developed a counterpart OF scenario in which, when a local fast-failover alternative port is not available, instead of forwarding back packets, a ``port down'' notification is sent at the controller, which in response enables the detour by updating the flow table at the reroute node. Figure \ref{fig:failure-13-12} shows the topology used for the experiments. In both OS and OF, the routing policy is the same and represents the optimal solution for the model presented in \cite{Cap15} when using as input the data of the ``Norway'' backbone instance (topology and traffic demands) obtained from \cite{SND}.

\begin{figure}
   \centering
   \subfloat[][]{
     \resizebox{0.5\columnwidth}{!}{
       \begin{tikzpicture}
         \begin{axis}[
				xmin = 0, xmax = 210,
				ymin = 0, ymax = 80,
				ylabel={Number of lost packets},
				ylabel style={yshift=-0.5em},
				legend cell align=left,
				legend style ={ at={(0.03,0.97)},anchor=north west, draw=black,fill=white,align=left},
				xlabel={Traffic rate $ [ packet/sec ] $}
			]

			\addplot[color=black, mark options={solid}, mark=*]
							coordinates{
						(20,8)
						(40, 21) 
						(60, 31)
						(80, 42)
						(100,52)
						(120,68)
						(140,68)
						(160,65)
						(180,69)
						(200,73)
						};
			\addlegendentry{OF 12ms}

			\addplot[color=black, mark options={solid}, mark=square*]	
							coordinates{
						(20,7)
						(40, 19) 
						(60, 28)
						(80, 35)
						(100,41)
						(120,59)
						(140,59)
						(160,60)
						(180,61)
						(200,58)
						};
			\addlegendentry{OF 6ms}

			\addplot[color=black,mark options={solid}, mark=triangle*, mark size=3]	
							coordinates{
						(20,3)
						(40, 17) 
						(60, 25)
						(80, 34)
						(100,38)
						(120,54)
						(140,51)
						(160,54)
						(180,52)
						(200,56)
						};
			\addlegendentry{OF 3ms}

			\addplot[color=black, mark=diamond*, mark size=3]
							coordinates{
						(20,1)
						(40, 14) 
						(60, 20)
						(80, 28)
						(100,33)
						(120,43)
						(140,46)
						(160,47)
						(180,46)
						(200,45)
						};
			\addlegendentry{OF 0ms}

			\addplot[solid, every mark/.append style={solid, fill=black}, mark=x, mark size=3]
					coordinates{
						(20,0)
						(40,9) 
						(60, 13)
						(80, 15)
						(100,19)
						(120,24)
						(140,25)
						(160,23)
						(180,27)
						(200,25)
					};
			\addlegendentry{OS}

			\end{axis}
       \end{tikzpicture}
     }
     \label{fig:ft-chart-losses}%
   }
   \subfloat[][]{
     \resizebox{0.5\columnwidth}{!}{
       \begin{tikzpicture}
			\begin{axis}[
                xmin = 0, xmax = 210,
                ymin = 0, ymax = 80,
                ylabel={Recovery delay $ [ ms ] $},
                ylabel style={yshift=-0.5em},
                legend cell align=left,
                legend style ={ at={(0.03,0.97)},anchor=north west, draw=black,fill=white,align=left},
                xlabel={Traffic rate $ [ packet/sec ] $}
            ]

            \addplot [color=black, mark options={solid}, mark=*,smooth]
            coordinates{
                (20,31.54525)
                (40,38.85555) 
                (60, 44.57435)
                (80, 52.30425)
                (100,58.80475)
                (120,65.3303)
                (140,68.48205)
                (160,68.49595)
                (180,69.42015)
                (200,66.85205)
            };
            \addlegendentry{12ms}

            \addplot[color=black, mark options={solid}, mark=square*,smooth]	
            coordinates{
                (20,22.16625)
                (40,23.645) 
                (60, 23.55755)
                (80, 25.6817)
                (100,29.35445)
                (120,33.55895)
                (140,33.54015)
                (160,34.27633)
                (180,36.3087)
                (200,35.3906)
            };
            \addlegendentry{6ms}

            \addplot[color=black,mark options={solid}, mark=triangle*, mark size=3,smooth]	
            coordinates{
                (20,15.0983)
                (40,13.9631) 
                (60, 14.81832)
                (80, 15.2102)
                (100,18.96975)
                (120,24.68516)
                (140,21.92705)
                (160,23.17205)
                (180,25.6405)
                (200,23.33795)
            };
            \addlegendentry{3ms}

            \addplot[color=black, mark=diamond*, mark size=3, smooth]
            coordinates{
                (20,7.621)
                (40, 7.872667) 
                (60, 8.829)
                (80, 9.52305)
                (100,11.969)
                (120,16.37455)
                (140,16.621)
                (160,15.85785)
                (180,18.75605)
                (200,18.53347)
                };
   			\addlegendentry{0ms}

            \end{axis}
       \end{tikzpicture}
     }
     \label{fig:ft-chart-delay}%
   }
  \caption{Failure recovery results: (a) lost packets and (b) recovery delay in OF}
  \label{fig:ft-charts}
\end{figure}
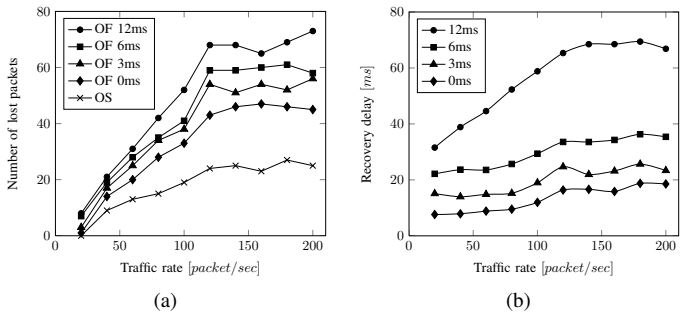

Figure \ref{fig:ft-charts} shows the number of lost packets caused by a failure of link (11,26). We generated traffic for 9 demands, each one having a non-local detour path for this specific failure, as for demand (22,10) in Fig.~\ref{fig:failure-13-12}. The experiment has been carried out by considering an increasing traffic rate from 20 to $200pkts/s$ for each demand, and 4 different values of 0, 3, 6, and $12ms$ of switch-controller RTT. Both in OS and OF, the fast-failover group type is used to detect the failure, hence we can assume equals detection delay. In Fig.~\ref{fig:ft-chart-losses} the higher losses in OF are due to the detection delay plus the recovery delay introduced by the controller (packets are dropped while waiting for the controller reaction), indeed losses increase accordingly to the switch-controller RTT. On the other hand, losses in OS are smaller because of the dependency only on the detection delay (packets are bounced back). Since the controller is not involved, the OS curve is not influenced by the switch-controller RTT. Figure \ref{fig:ft-chart-delay} depicts the recovery delay interval in OF between the sending of a ``Port Status'' notification to the controller and the update of flow tables. This results does not apply to OS as no packets are dropped and hence we assume the recovery is instantaneous.

\section{Conclusion}
\label{sec:conclusion}

Using a stateful data plane in SDN allows to delegate control tasks to switches with significant gain in performances and scalability of traffic management applications, along with reduced complexity at the controller. OpenState is an example of a data plane abstraction that allows to process packets in a stateful fashion on the basis of packet-level events and timers. We presented here two applications, namely forwarding consistency and failure recovery, that greatly benefit from a stateful SDN data plane. In the forwarding consistency case, we argue that programmers should be able to define the granularity and lifetime of forwarding decisions, while in the failure recovery case we shown how simple (just a tag) switch-to-switch signaling allows to instantaneously react to distant failures. We formally described the data plane behavioral model of both applications in the form of a Mealy machine. Experimental results have been provided showing the advantages of an OpenState-based implementation in terms of processing delay and number of lost packets.

\section*{Acknowledgment}
This work has been partly funded by the EU in the context of the ``BEBA'' project \cite{beba} (Grant Agreement: 644122).

\bibliographystyle{IEEEtran}
\bibliography{biblio}

\end{document}